\documentclass[journal]{IEEEtran}

\usepackage{makeidx}
\usepackage{graphicx}
\usepackage{multirow}
\usepackage{multicol}
\usepackage{rotating}
\usepackage{cite}
\usepackage{caption}
\usepackage{subcaption}
\usepackage{multirow}
\usepackage{amssymb}
\usepackage{epsfig}
\usepackage{epstopdf}
\usepackage{float}
\usepackage{balance}
\usepackage{booktabs} % For Table
\usepackage{url}
\usepackage{amsmath} % For Math
\usepackage{tikz}

\usepackage[ruled,vlined]{algorithm2e}
\usepackage{algpseudocode}
\usepackage{pifont}
\usepackage{tikz}
\usepackage{lscape}
\usepackage{multirow}
\usepackage{longtable}

\usepackage{hhline}
\ifCLASSINFOpdf

\else
 
\fi

\begin{document}

%% Title, authors ad addresses

\title{A First Look at Privacy Analysis of COVID-19 Contact Tracing Mobile Applications}
\author{Muhammad Ajmal Azad, Junaid Arshad, Syed Muhammad Ali Akmal, Farhan Riaz,Sidrah Abdullah, Muhammad Imran and Farhan Ahmad \\
%\color{red} {Farhan Ahmad and Sidrah made no contribution despite asking.}
\IEEEcompsocitemizethanks{\IEEEcompsocthanksitem 
Muhammad Ajmal Azad and Farhan Ahmed are with Derby University, Derby, United Kingdom, Junaid Arshad is with Birmingham City University, Birmingham, United Kingdom, Syed Muhammad Ali Akmal, and Sidrah are with NED University of Engineering and Technology, Karachi, Pakistan, Muhammad Imran is with King Saud University, Saudi Arabia and Farhan Riaz is with National University of Science and Technology, Islamabad, Pakistan}}

\maketitle

\begin{abstract}
Today’s smartphones are equipped with a large number of powerful value-added sensors and features such as a low power Bluetooth sensor, powerful embedded sensors such as the digital compass, accelerometer, GPS sensors, Wi-Fi capabilities, microphone, humidity sensors, health tracking sensors, and a camera, etc. These value-added sensors have revolutionized the lives of the human being in many ways such, as tracking the health of the patients and movement of doctors, tracking employees movement in large manufacturing units, and monitoring the environment, etc. These embedded sensors could also be used for large-scale personal, group, and community sensing applications especially tracing the spread of certain diseases.  Governments and regulators are turning to use these features to trace the people thought to have symptoms of certain diseases or virus e.g. COVID-19. The outbreak of COVID-19 in December 2019, has seen a surge of the mobile applications for tracing, tracking and isolating the persons showing COVID-19 symptoms to limit the spread of disease to the larger community. The use of embedded sensors could disclose private information of the users thus potentially bring threat to the privacy and security of users. In this paper, we analyzed a large set of smartphone applications that have been designed to contain the spread of the COVID-19 virus and bring the people back to normal life. Specifically, we have analyzed what type of permission these smartphone apps require, whether these permissions are necessary for the track and trace, how data from the user devices is transported to the analytic center, and analyzing the security measures these apps have deployed to ensure the privacy and security of users.
\end{abstract}


%\linenumbers

\section{Introduction}
\label{S:1}
It is believed that a virus that causes the novel COVID-19 disease spreads mainly from having a close interaction or contact with the person already being affected with the virus, and still carrying the attributes of the virus. Since December 2019, work has already begun on the development of a potential vaccine for the cure, however, until the development of vaccines for masses, the only possible way of protection is to limit the interaction with the people, isolate the people through imposing full or smart lock-down. The efficient smart lockdown could be imposed by employing the track and trace method that only isolates the infected people and their contacts.  For this purpose, several smartphone apps for android, iOS, and Windows operating systems have been developed for tracing, tracking, and informing citizens about whether they have recently come in close contact with the person showing confirmed attributes of COVID-19. These apps can be either private or government-owned.

\begin{figure*}[h!]
\centering
\includegraphics[width=15cm,height=8cm]{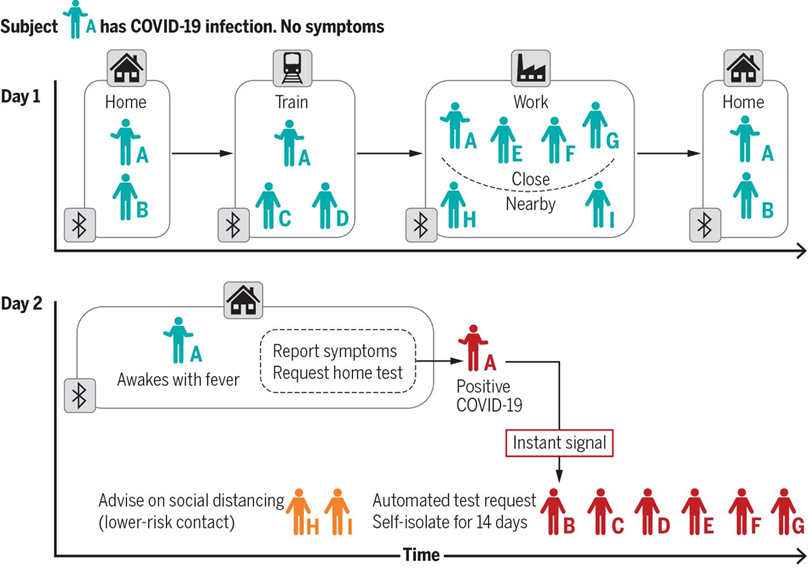}
\caption{Tracing and Isolating through Smartphone based Contact-tracing \cite{Ferrettieabb6936}}
\label{fig:tracing}
\end{figure*}

\color{blue} The Contact tracing operates by identifying positive cases of COVID-19 and asking them for their close contacts manually or identifying their close contacts in an automated way.  Figure \ref{fig:tracing} presents the building block and working mechanism of the contact tracing apps \cite{Ferrettieabb6936}.
The contact-tracing apps exchange information when the phones of two persons are close enough to each other. These people who came socially close to one another will be informed if their counterparts during the social interaction were officially infected with the COVID-19. These apps can only serve as the anchor point to inform citizens and provide suggestions on whether they should go to isolation or not. These apps have already shown efficiency in controlling the spread of the virus in South Korea and Singapore and flatten the spread curve through "Test" and then  “Trace” mechanism \cite{southkorea,southkore1}.
\color{black}

The contact-tracing apps are broadly developed by the national or country lead health regulators. To provide a reliable and efficient decision, the developed apps utilize the information from various smartphone sensors (GPS, Bluetooth) along with names, addresses, gender, age, contact details, calling log history, and contact history, etc. to make the decision. These apps either interact automatically with the national health data system for the test results of the citizens or citizens could manually provide test results to the health organization. For instance, upon downloading, Pakistan's track \& trace app \footnote{https://tinyurl.com/y4lt634d} requests permission to use the device's location and user personal details such as name, phone number, and email address, and that stored data will be shared with a third party. The current version of this app (as of June-2020) then uses the device's location and the location of users with positive COVID-19 test results to render a map showing high-level \textit{hotspots} for COVID-19 infections. Similarly, Google and Apple jointly developed an API that enables app developers to use the Bluetooth beacon messages for exchanging information between two persons who are in close contact with each other and showing virus symptoms. Some countries use the call detailed records to trace the close contact of the infected person and isolate them as well. 

The use of these apps is normally voluntary and is considered as the support to control the spread of the virus. The developed app requests users for specific permission e.g. contact details, call history, web searches, camera permissions, access to call records, messages, and mobile media (videos and photos). This information could pose serious privacy and security risks to the users and limit users to use these apps. The privacy of users may be protected through the use of different mechanism e.g. data anonymization, differential privacy, and decentralized app development \cite{Troncoso2020}. However, it is already identified that anonymization systems are not providing effective privacy-preservation \cite{ga1,ga2,ga3} and decentralized app development is still at the early stage and progressing very slowly.

In this paper, we provide a first look at the permission, security, and privacy analysis of the contact tracing apps available at the Android and iOS app stores. We have studied app stores to define the nature of the privacy risks these apps have, what types of permissions these apps are requesting for their functioning, which permissions are unnecessarily required, and how they store and process the user data. Currently, only a small number of certified apps (developed by the country regulator) are available at the play stores, so we performed an exhaustive manual analysis on the available apps. Our analysis shows that the majority of the track and trace apps collect personal information such as name, device ID, and location, however, some apps require access to further resources such as SMS, microphone, camera, and storage memory of the device. Access to such resources is not required for the accurate function of such apps, and, therefore should not be requested by the developers. Furthermore, a number of apps disclose sharing data with third parties, however, a small number of these acquire the permission of the user before sharing this data with the third parties. As track and trace apps are voluntary and rely on the public's trust to achieve their function effectively \cite{IND2020}, addressing concerns with regards to data collection and sharing are paramount to their success to combat COVID-19.  

The rest of the paper is organized as follows: Section \ref{sec:related} provides a discussion on privacy and permission analysis of mobile apps and work performed towards the development of track and trace apps. Section \ref{sec:problem} defines the background on the trace and track smart applications. The section also provides the working mechanism and important features of developed apps. Section \ref{sec:analysis} critically analyses the security and privacy of different apps. Section \ref{sec:conclusion} provides recommendations for design chose secure development and concludes the paper. 

\section{Related Work}
\label{sec:related}

A large number of works have been presented that analyse the functionality of smartphone apps and the leakage of the sensitive information of their users \cite{S1,S2,S3}. Many contact tracing applications involve tracing users using GPS, Bluetooth, and wireless technologies \cite{ct12, ct13, ct14, ct15, ct16, ct17}. These approaches usually provide users with two options. Either the user has to self-report themselves, or the application takes the help of a wireless technology \cite{ct11}. A large number of people are currently downloading and using contact-tracing apps, and hence the privacy aspects of these apps have become paramount for the research and development of unanimous privacy regulation. The regulators such as the Federal Trade Commission, the US National Telecommunication and information administrations, the European Union Commission \cite{eu1,eu2}, Information Commissioner’s Office \cite{ICO} are analyzing and providing these guidelines to the app developers, content creators, website operators to improve the development of their products in terms of security and privacy. In this section, we summarize works related to the security and privacy of smartphone apps. 

Smartphone apps normally get access to user data and other information through the use of permission \cite{appper} that the users provide to the smartphone app at the time of app installation. For example, an app might ask to grant the to see the location of the user, the messages stored on the mobile phone, the search history, etc. The user can still control the permission after the installation but it might affect the functionality of apps operations. Providing permission to various private information would expose the private information of users to the advertisers, insurance companies \cite{per1,per4}, and publicly expose personal data of the users without the user’s consent \cite{per2,per3}. A large number of smartphones app also ask unnecessary permission that is not required for the functionality of the app, these apps might pose a serious threat to privacy and security of the users \cite{un1,un2}. Muhammad et al. \cite{muhammad,muhammad1} analyzed the security and privacy of smartphone apps designed for blocking the advertisement and providing mobile VPN clients. Ilaria et al. \cite{Liccardi} analyzed the permissions requested by the smartphone apps and assigned a sensitivity score to the app if the app asked to read the personal information of the users. They concluded that around 56\% of the app asks users to provide permission to sensitive parts of the user’s data. Barrera et al. \cite{berral} investigated the relationship between free android apps and the most popular 1100 Android apps by deploying machine learning methods. It is also concluded that people are willing to use the paid version of smartphone apps if apps are not asking for unnecessary permissions. Enck et al. \cite{enck} proposed a lightweight certification mechanism to identify Android apps that are asking for suspicious permissions. Pern et al. \cite{pern} studied the user-consent permission systems by using the user-centric data from the Facebook apps, chrome browser extensions, and Android smartphone apps. It is very important to develop tools or applications that inform users about the privacy indicator of the apps they are using for specific purposes. To address this issue, Max et al. \cite{max} developed a prototype that provides users with privacy indicators of the app. The prototype also identifies previously exposed hidden information flows out of the apps.

Contact tracing with smartphones can be employed to restrict the transmission of a pandemic disease.  Utilizing computing technologies to avert and control the pandemic seems to be an obvious choice.  However, these contact tracing apps might invade privacy, collect personal data, and justify mass surveillance against users' wishes. There must be a protocol for contact tracing that observes commitment to privacy, as well as provides the consent mechanism where there is a need to share individual data. Contact tracing may collect personal data such as location which is not an effective privacy control when it comes to user's data \cite{ct7}. The process of contact tracing usually involves collecting users' privacy information without informing them. Privacy-literate individuals might be reluctant to share their information which in turn hampers the process of contact tracing. Privacy-preserving approaches might encourage individuals to participate more in this process and increase their confidence in those applications \cite{ct11}. 

Prominent privacy researchers from across the world are arguing with the government agencies and vendors involved in developing the contact-tracing application about the privacy, and also highlighted the catastrophic consequences these apps would have on the citizen’s private lives \cite{ross2,ross1}. To ensure privacy, Berke et al. \cite{Cho2020ContactTM} utilized the semantics of private set interaction for assessing the risk exposure of users using encrypted and anonymous GPS locations. Manish et al. \cite{manish} analyzed the privacy preservation mechanism for various contact tracing applications and discussed the attributes which contact-tracing apps should have to ensure the privacy of users.\color{blue} Michael et al \cite{ethics} discussed the ethical consideration of contact-tracing apps for fighting against the COVID-19. Several contact-tracing application has been compared in \cite{comp} in terms of data collection, retention of data, purpose, and sharing of collected data, what mechanisms the apps have deployed to ensure the privacy of users. Most recently, Carmela et al. \cite{Troncoso2020} describe and analyse a decentralized system for secure and privacy-preserving proximity tracing to combat the spread of COVID-19. The system is solely based on the anonymous identifiers of positive users of the COVID-19 without providing the exact location information to the health authorities. Serge \cite{cryptoeprint} analyse the security and privacy properties of the pan-European Decentralized Privacy-Preserving Proximity Tracing (DP3T) system. Health authorities or any other users would not be able to learn the private information of the users except a notification message when a person is exposed to COVID-19 affected person.Ruoxi et al. \cite{sun2020vetting} analyze the security and privacy of contact tracing apps in three dimensions: a) evaluate the design choice (centralized or decentralized) used for privacy preservation, b) static analysis for the identification of potential vulnerabilities and c) (iii) evaluate the robustness of approaches used for privacy-preservation. The paper has not analyzed the permission analysis. Yaron et al. \cite{Yaron} analysed the security and privacy properties of Bluetooth based specification by Apple and Google, concluding that the specifications may have some significant security and privacy risks.  The Centers for Disease Control and Prevention (CDS) have issued guidelines that define a set of features a contact tracing app should have to help health departments to overcome the COVID-19 pandemic \cite{cdc}. 
\color{black}

\section{CHARACTERIZING Contact Tracing Apps}
\label{sec:problem}
Contact tracing is an important tool for the community to prevent the outspread of novel pandemic diseases, such as COVID-19 \cite{ct1}. In past contact-tracing tools have shown effectiveness against the spread of transmissible diseases such as STD, HIV, Ebola, and tuberculosis \cite{ct9, ct10}. %It refers to the individual-level spread of information and the network of likely transmission routes of a disease \cite{ct2}. 
Contact tracing is the process of identifying persons who are in close contact with the infected person so that exposed targets can be informed to have self-isolation and quarantine, thus breaking the chain of transmission \cite{ct7}. The current outbreak of COVID-19 and its highly contagious feature motivates technology developers to develop smartphone apps for the effective tracing of the footprint of the disease. In this section, we provide the architectural setup of contact-tracing apps and their significance towards controlling the spread of disease.

\subsection{Centralized and Decentralized Architecture}
\begin{figure*}[h!]
\centering
\includegraphics[width=15cm,height=5cm]{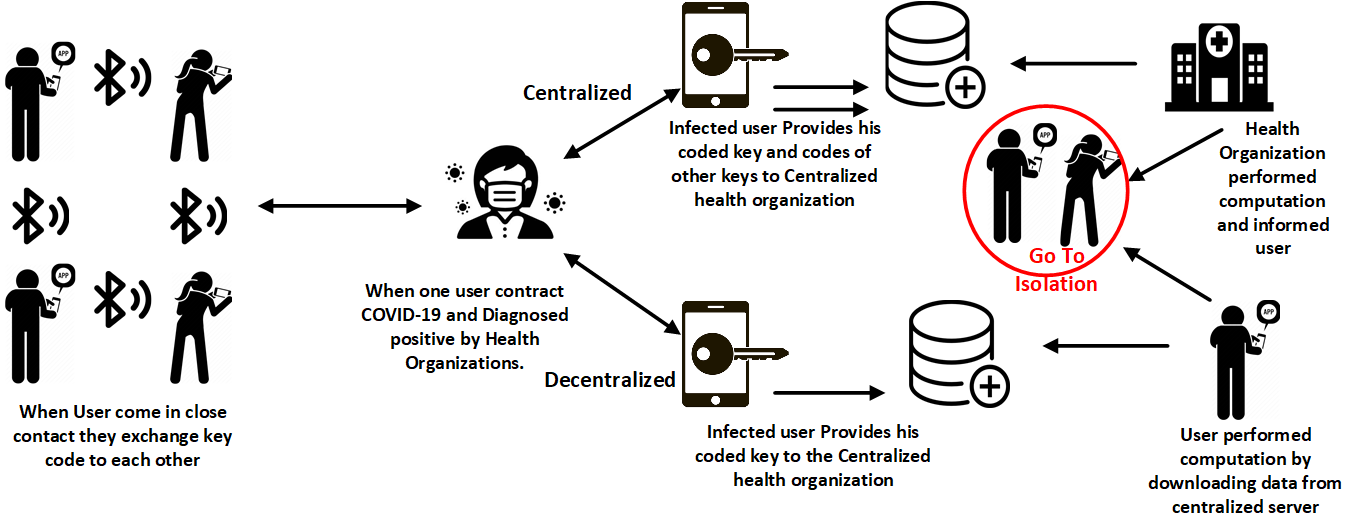}
\caption{Architectural Setup of Contact Tracing Apps}
\label{fig:cendcen}
\vspace{-0.5cm}
\end{figure*}

\color{blue} The design of contact-tracing apps is mainly using data from the users thus has some privacy concerns which motivate the developer to come up with privacy-preserving solutions. The privacy of users can be addressed using the centralized and decentralized system setup. The centralized and decentralized apps entirely have different architecture and properties shown in Figure \ref{fig:cendcen} and explained below.  

\subsubsection{Centralised models} In the centralized setup, the smartphones of the users having specific contact-tracing apps send the random identifier to the centralized trusted system. The centralized system in this setup holds the information from all users of the app. If a person has tested positive for the COVID-19 virus, the identifier of other users who have exchanged identifiers in the past can be sent to the centralized server along with other information e.g. time data is sent, a time when identifiers are exchanged, etc. The centralized system decrypts the identifiers and automatically notifies the interacted phones suggesting or informing users to self-isolate or take other preventive measures. The centralized system can also utilize the available information for further analysis and policies for placing lockdown in hard-hit proximities.
 
\subsubsection{Decentralised models} In the decentralized setup there is not a trusted centralized system that exists for the handling of the user's data and matching of smartphone’s identifier. If a person is diagnosed positive with the COVID-19, the identifier of his phone and test result is uploaded to the centralized system. Other smartphones having the app can access these reports and locally establish the truth whether he was close to an infected individual or not. If a smartphone comes across the identity that has COVID-19 then alert is the sent to the user of the smartphone for precaution and self-isolation. The location and proximity of the person are not known to the centralized system thus ensuring the privacy of the users using the app.  The health organizations or the government still used the shared data to understand the spread of the virus in the community but would not have detailed information about the users. 
\color{black}
\color{blue}
\subsection{Significance of Contact Tracing for COVID-19}
Since the outbreak of the COVID-19 pandemic in December 2019, as of June 2020, there exists no medicine or vaccine to fight against the rapidly spreading pandemic. Governments across the world are currently focusing on the ways that would have the least load on their health systems. This has been achieved through imposing travel restrictions or lockdown however, it is not only affecting the economy but there are also fears of the second wave of infection once the restrictions are relaxed. The governments are finding ways to identify the methods for contact tracing in order to quickly identify and isolate the infected persons. The manual contact tracing is not only slow and has a late response, but would also require resources for identifying infected persons and then asking for his contacts and then contacts of his contacts to track the flow of the disease. The technologies soon realized the importance of smartphones and used the inbuilt smartphone sensors for tracking in an automated and efficient way.  The use of digital technologies help the citizen at the early stage of the virus spread and inform people for isolation at the early stage. The use of smartphone apps for contact tracing has shown promising results in several countries to combat the spread of the virus \cite{RePEc}, however, the performance efficiency depends on the number of people using the application \cite{tech}. One thing that limits the usage of the app is the privacy because a large number of existing apps store data at the central trusted system, and in some circumstance, this data is made available to the third party systems for performing artificial Intelligence and data analysis.
\color{black}
 \subsection{Vendor Support}
\color{blue}As the healthcare officers and medical entities are working together worldwide to fight the spread of this pandemic, Google and Apple have joined an effort and developed a privacy-preserving contact tracing API that uses Bluetooth signals \cite{ct18} for exchanging information between people who are in close contact with each other. The apps using this API operate in a decentralized fashion, however, a centralized database is maintained.
\color{black}
This framework allows healthcare agencies to propose or develop smartphone apps that help in limiting the spread of the disease with the help of Bluetooth technology. This API will bring interoperability between iOS and Android devices while maintaining privacy, consent, and transparency \cite{ct19}. A test project, PACT (Private Automated Contact Tracing), was built at MIT to harness the strength of Bluetooth-based, privacy-preserving, automated contact tracing API. This project detects proximity between contacts with the help of Bluetooth signals within a 6-foot radius. Instead of relying on the GPS, this system sends out random Bluetooth numbers, which can later be updated to a database with the user's consent \cite{ct21}. The first large-scale pilot for this joint venture has been launched in Switzerland, known as SwissCovid. This application determines the close contact that lasted for more than 15 minutes and notifies the user with the procedure to follow \cite{ct20}. 

Apple also released a new application for COVID-19 based on CDC guidelines that provide COVID-19 information across the USA. In this application, the users have to answer some questions related to recent exposure and risk factors. In return, they get a CDC recommendation on what their next step should be. However, this application does not replace a healthcare worker in any way \cite{ct29, ct30}. Another application, "HEALTHLYNKED COVID-19 Tracker", which became the most downloaded coronavirus tracker application for March. The application enables users to track local cases and chat with other users around the world. The most unique feature of this application is that it enables real-time chat with other users and share updates \cite{ct40, ct41}.

\section{ANALYSIS OF Contact Tracing apps }
\label{sec:analysis}
In this section, we present our approach to studying current contact tracing apps for COVID-19. We have focused on smartphone apps for any platform (iOS, Android, Microsoft) and available anywhere in the world. Although there appears a concerted effort by governments across the globe to contain the pandemic, we identified through our analysis that many such apps have been developed by third-party individuals or organizations. Therefore, we have also included these in our analysis. Furthermore, as the contact tracing technology is still in its early stages (especially within the context of COVID-19), although many apps claim to perform track and trace function, their effectiveness in this respect is subjective.  

\begin{figure*}[h!]
\centering
\includegraphics[width=18cm, height=4.6cm]{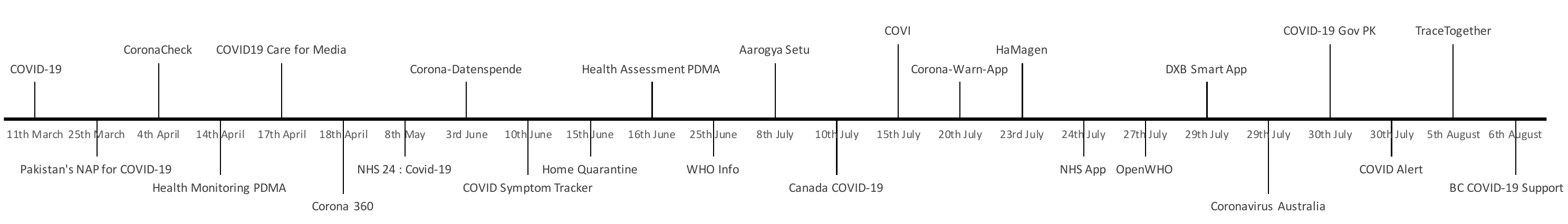}
\caption{Timeline for release of COVID-19 track \& trace apps}
\label{fig:timeline}
\vspace{-2.0em}

\end{figure*}

\subsection{Data collection}
\color{blue}
In order to achieve an in-depth analysis of current track \& trace apps, we performed exhaustive search techniques to collect relevant apps across two major platforms (iOS and Android), irrespective of the country or the developing organization (government and private). We have chosen iOS and Android because these systems currently hold the most market share in the smartphone industry. Specifically, we used keywords such as \textit{COVID-19}, \textit{COVID track \& trace}, and \textit{Coronavirus track \& trace}. We analyzed the results of our search queries to filter apps that did not relate to contact tracing to mitigate the spread of COVID-19. We used a manual analysis of the app description to conduct this filtering. Furthermore, as several countries have encountered difficulties to achieve effective contact tracing, we did not exclude apps performing the partial or limited function in this regard. Overall, we identified 26 smartphone apps that claim to perform contact tracing in their description belonging to 17 different countries. Details of these apps are presented in table \ref{tab:analysis} with a brief description of some apps presented below. Furthermore, Fig \ref{fig:timeline} presents a graphical representation of the timeline with respect to launching dates of prominent apps. 

\color{black}
\begin{itemize}
   \item \textit{COVID-19 Gov PK} is an app developed by the government of Pakistan. Initially, the application provided awareness to citizens about COVID-19, however, with the development of the new \textit{radius alert} feature, this application provides information which are hot-spot areas that help country to impose smart lock-down \cite{ct22}. 
   \item \textit{Health Canada} is developed by the government of Canada to provide a personalized recommendation to the users based on their risk factors. Personal data collected is only used by Health Canada and is not shared with any other application or agency \cite{ct31}. 
    \item The government of Vietnam has developed an application named \textit{COVID-19} which includes features such as chatbot, consultation, and live updates on COVID-19. The application requires access to media, location, storage, device ID, and call logs. The application's privacy policies are updated in its native language \cite{ct32}.
    \item  \textit{COVID19 - DXB Smart App} is developed by the government of Dubai and provides general information on COVID-19 and also provides correct statistics. The application collects personal information voluntarily but does not share with the third-party applications unless required by the law \cite{ct34}. 
     \item \textit{COVI} is a third-party COVID-19 informative app developed by Droobi, a Qatar based digital company. This application collects personal data such as contact information, age, health information, and unique identifiers, etc. \cite{ct33}.
    \item \textit{Corona360} is an app developed in South Korea which enables users to update their COVID-19 status as well as view the status of other people. For privacy reasons, the app does not collect any personal information such as ID, name, or phone number \cite{ct35}. 
    \item \textit{CoronaCheck} is a third-party application that has been developed to enable its users to conduct self-assessment and provide accurate expert COVID-19 information to the users. This application does not collect any personal information and does not share data with third-party vendors \cite{ct36}. 
    \item \textit{The Beat COVID Gibraltar} app is developed for the region of Gibralter and it utilizes Bluetooth technology in a decentralized way to track other phones who come in close contact with the person declared himself as the virus affected.

\item \textit{BC COVID-19} Support is developed for the resident of British Columbia, Canada, to inform the people about the status of COVID-19 in British Columbia and guide them on what next action people should take. All the recommendations are personalized so it involves some level of contact tracing.

\item \textit{COVID Symptom Study} has been designed for everyone to report their health status to the people who are developing policies to fight against the Virus. 

\item \textit{BeAware Bahrain} is a mobile application developed for the region of Bahrain that helps citizens to contain the spread of the virus by using the contact tracing efforts.

\item \textit{Tawakkalna (Covid-19 KSA)} is the official app of kingdom of Saudi Arabia. It helps in controlling the spread of COVID-19 and suggest authorities where to impose the curfew and lockdown.

\end{itemize}

Our study of these apps consisted of analyzing publicly available information shared by the app developers and platform i.e. privacy policy, permissions requested, and user reviews. Furthermore, as some of the apps did not use SSL, we performed black-box testing of the apps using the Burp suite to analyze the network traffic during the app usage. The traffic analysis did indeed help us identify the information collected by these apps and shared with back-end servers which are liable to interception using network sniffing software.  

\begin{figure}[h!]
\centering
\includegraphics[width=9cm,height=5.8cm]{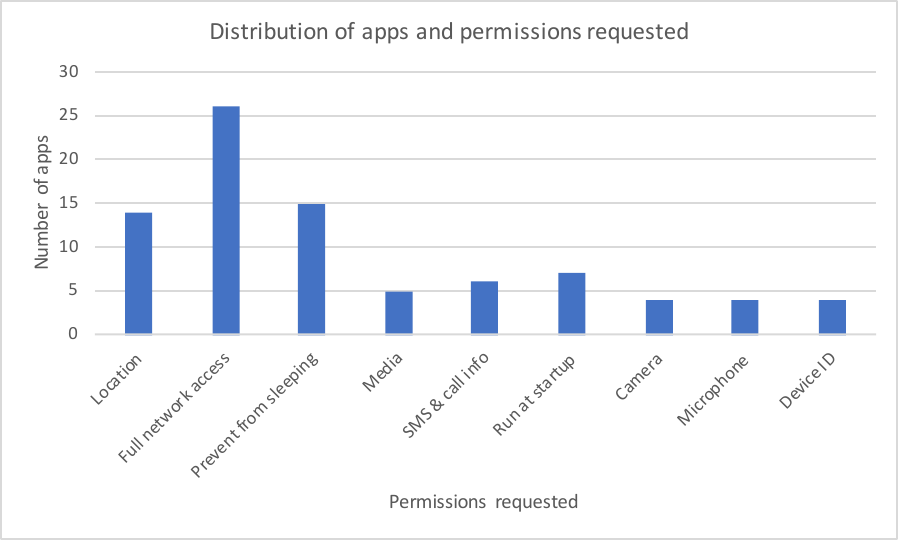}
\caption{Distribution of apps and requested permissions}
\label{fig:permissions_analysis}
\vspace{-2.0em}

\end{figure}

\subsection{Permission analysis}
Permissions required by a smartphone app are significant as they communicate with the user the resources required by the app to perform its function. Therefore, presenting the user with a list of permissions not only achieves transparency (providing the user an insight into the app operation) but it also serves to seek user consent. Within the context of our study, we have gathered information about the permissions required by the track and trace apps under study. Figure \ref{fig:permissions_analysis} presents a graphical representation of the distribution of the apps with respect to permissions required by them. 

As presented in Figure \ref{fig:permissions_analysis}, we expected the majority of the apps to require access to location data of a device however our study also identified permissions requested by the apps which are not necessary to perform their function. For instance, we identified 06 apps that require access to SMS and call information of the device. In some scenarios access to phone numbers can be envisioned however access to SMS within a device is not essential to the function of a track and trace app. Similarly, we identified 04 apps that require access to the camera of the device which is of course an unexpected request by a track and trace app. 

Furthermore, 04 apps studied require access to the microphone of the device whereas 05 apps require access to media and storage of the device which are of course not critical to the app's function. Such apps are indeed causing concern regarding the privacy of their users and exemplify a lack of attention to the security and privacy of the users by app developers.

%\begin{figure}[h!]
%\centering
%\includegraphics[width=6.4cm,height=5.8cm]{images/covid-SSL.pdf}
%\caption{Distribution of apps with respect to SSL support}
%\label{fig:ssl}
%\vspace{-2.0em}

%\end{figure}

%\begin{figure}[h!]
%\centering
%\includegraphics[width=7cm,height=5.8cm]{images/apps-platform.pdf}
%\caption{Distribution of apps with respect to platform support}
%\label{fig:platforms}
%\vspace{-2.0em}

%\end{figure}

\subsection{Analysing privacy}
To understand the privacy considerations applied for the apps within our study, our analysis took into account the privacy policy published by the application developer as well as the use of basic privacy protection mechanisms such as SSL/TLS to achieve encrypted data transmissions. 

\color{blue}
Through the study of privacy policies of the apps as well as traffic analysis, we identified that most of the apps collect personal data such as location information, name, and phone number, etc. Although collecting such information is vital for effective track \& trace, appropriate mechanisms should be applied to ensure secure sharing, processing, and storage of such data. Such details were not available for most of the apps analyzed in this study. Furthermore, an interesting observation we made was the type of information gathered by the apps. Specifically, our analysis identified that the track \& trace app developed by Dubai collects personal information such as date of birth, name, email address, and caller ID. Such information is not required for effective track and trace and risks privacy of users as any malicious actor with access to such information can easily perform ID theft attacks. We believe data collection policies for such apps require immediate attention to minimize risk to individual user privacy. 

A significant challenge with respect to security and privacy within smartphone apps is sharing data with third parties to aid targeted advertisements. Smartphone apps using such strategies leverage advanced analytics techniques to identify user-behavior and profiles to achieve personalized advertisements. Within this context, the information collected by track \& trace apps is highly personalised and if made accessible to third-parties can lead to sophisticated advertisement techniques breaching public trust and confidence in such apps. Our analysis of the apps concluded that all apps are free to use and do not include any in-app advertisements. However, some apps (as highlighted in table \ref{tab:analysis}) do share data with third parties without precise information about who these third parties, what data is shared, and how this data will be used by the third parties. Therefore, focused efforts from the research community are required to enhance data sharing, processing, and storage practices within such scenarios.

Another aspect concerning the privacy of information collected is how it is stored and shared by the apps. In this respect, our analysis revealed that 10 of the apps shared data with \textit{third parties} however the nature and identity of such parties are not identified in the privacy policies. This is a cause of concern regarding individual privacy as the aims of sharing this data are not clear and therefore users are not aware of how their data may be used. For instance, a common apprehension among users is the sharing of data with advertisement agencies who may wish to use such data for targeted advertisement and adware. 

\color{black}
Having said this, we also identified examples of good practice within our analysis. Specifically, some of the apps clearly state requirements for user consent before sharing data with third parties thereby assuring users with respect to how their data is shared. For instance, Corona360 (the app developed by the government of South Korea) collects personal and sensitive data of users but whenever the data is used, the user is notified for the reason of data usage. 

In addition to the above, our analysis also uncovered security vulnerabilities within some apps. In particular, we identified five apps that were not using SSL/TLS to ensure secure communication made through the app. Pursuing this direction of analysis, we conducted traffic monitoring of such apps and identified serious flaws in the app developed by the government of Pakistan. Details of these vulnerabilities have been reported to the relevant authorities however such vulnerabilities do put user privacy at risk especially where the app is collecting and utilizing personal user data. 

\subsection{App review analysis}
In addition to the permissions requested, privacy policy, and traffic analysis of the worldwide COVID-19 track and trace apps, we have also studied user comments available in Google Play and App Store reviewing these apps. Although the majority of these comments are related to the usability and general function of the app, we found some comments to be insightful with respect to how the app collects and utilizes data. For instance, the app developed by the government of Israel was commented by a user to ask for users' permission when sharing data with third parties as well as to \textit{guarantee not to send the data anywhere but compare it locally on user's device against downloaded "Corona paths"}. Another interesting observation was made for the Corona-Datenspende app, where user comments suggest that \textit{one cannot use the app without connecting to a fitness account and hence completely breaking the point of anonymity.} As suggested by the user, connecting to the app via a fitness app such as Fitbit indeed does indicate sharing personal user data across different apps which is a risk to user privacy. 

Through analysis of the user reviews of apps, we observed that although some of the users have included concerns about the privacy of information through their feedback, these are relatively minor proportions of users. For instance, for the \textit{COVID-19 Gov PK} app, user review includes comments highlighting a lack of encryption and concerns about data traveling in plain-text. However, for apps such as \textit{Corona-Datenspende} which requires a user to connect to a fitness app as a pre-requisite, there are no user comments with regards to how data is captured, analyzed, stored and processed between the third-party fitness app and the government app for CVOID-19. These observations reflect a lack of awareness among users with regards to measures to preserve the privacy of personal data collected, stored, and analyzed by computing systems, therefore, requiring efforts to raise awareness among users.

\section{Conclusions and Recommendations}
\label{sec:conclusion}
%The outbreak of COVID-19 has put many countries in strict or partial lockdown to control the spread of the virus within the communities. This would help countries to control the load over their health systems. Governments across the world are investigating to find ways to bring economic activities back on track without risking the lives of their citizens and without overloading their health system. One way to achieve these objectives is to place the contact-tracing smartphone application that helps in tracing the contact of an infected person and suggesting them for self-isolation and quarantine. 
As Coronavirus is a contagious disease that spreads through close social interaction between humans, contact tracing is vital for containing its spread. Mobile devices present an ideal platform to introduce contact tracing software due to their ease of use, widespread ownership, and personalized usage. Therefore, several smartphone apps have been developed by governments, international agencies, and other parties to mitigate the virus spread.
%Over the past few months, the usage of contact-tracing apps continues to skyrocket, providing effectiveness towards the containment of the virus, 
However, there is an increasing concern regarding the collection and use of data, and out-sourcing data to third-party systems. In this paper, we analyzed a large set of contact-tracing apps with respect to different security and privacy metrics. Specifically, we analyzed contact-tracing apps for permission analysis, privacy analysis, the security of the apps, and reviews of the users. \color{blue} Our major findings are as follows:
\begin{enumerate}
    \item Although there have been significant technological advancements to aid COVID-19 response, contact tracing apps require further enhancements to achieve desired objectives in a privacy-aware manner.
    \item A number of track \& trace apps request permissions which may not be required for the successful operation of the app's function. These include access to storage media, camera, and microphone which might result in in a breach of user's privacy.
    \item Several apps mention outsourcing data to third parties, however, it is unclear who are these third parties, what data is shared, and how it is processed by these parties.
    \item Some apps (used in developing countries) have not adopted appropriate security measures for the exchange of the data to and from the user to the data centers.
    \item Our analysis of the user reviews and the ratings for contact-tracing apps suggested that a large number of users are aware of privacy concerns of these apps.
\end{enumerate}

Though digital technologies could play a prominent role in addressing the current pandemic challenges and the containment of the spread of the virus. However, the effectiveness and accuracy of these systems depend upon the working architecture of applications and user participation. The user participation could be improved if systems employed mechanism that ensures the security and privacy of users. To ensure the privacy, security, and secure development of contact tracing apps,  we recommend following design choices that should be followed for the development of contact tracing apps:

\begin{enumerate}

\item In order to ensure the privacy and security of the user data, the contact tracing systems have to consider the well established and state of the art encryption systems for storing data, enable personalized access control mechanisms and utilize secure communication mechanisms for the exchange of data between the users and the data center. Furthermore, developers should also consider the semantics of secure software development, strong authentication mechanism possibly two-factor authentication to minimize the risk of misuse.
\item 	The contact-tracing apps should perform their operations in a completely decentralized way. i.e. the system performs the bulk of its operation at the user side.
\item The app's privacy policy should be mentioned in a way that a user could easily understand. The developer should also adopt the mechanisms that they could easily destroy user data once this pandemic is over.
\item The design system should not unnecessarily seek permissions for example access to videos, browsing history, or the images.
\item The developers should consider the measures that assign a unique pseudonymized identifier for the users which must not be linked to the user's real identity and could not be used to learn the private information of users through background knowledge. 
\item To improve usability, the design should be simple and should have a user interface for interaction and personal tracking.
\item We also recommend developers and regulators to use the identity verification (telephone number authentication) or authentication system within their trace and track system so the information could be exchanged through the reliable voice call. 
\end{enumerate}

It is very important to incorporate the techniques that ensure the privacy of citizens so that they can confidently participate in limiting the spread of the disease. The apps should not serve as the tool for mass surveillance tools so that people trust the system without having any concerns about their privacy and tracking of their private lives. As a part of our future work, we are looking to conduct a user study using qualitative measures focused on directly considering the feedback from users to further understand the usability and security concerns of users.

\color{black}

\clearpage
\onecolumn

\fontsize{6.5}{7}\selectfont

\centering
\begin{longtable}{|p{1cm}|p{1cm}|p{2.8cm}|p{3.0cm}|p{0.8cm}|p{0.8cm}|p{0.5cm}|p{3.0cm}|p{0.7cm}|p{0.8cm}|}
\caption{Analysis of Smartphone apps designed to limit spread of COVID-19.}
\\\hline 
 \textbf{App} & \textbf{Platform} & \textbf{Permissions Requested} & \textbf{ Privacy Policy} & \textbf{Country}& \textbf{No of Downloads}& \textbf{TLS/ SSL} &  \textbf{App Reviews} & \textbf{App version} &\textbf{API version}\\ \hline

COVID-19 Gov PK  & Android & Location (approximate and precise), full network access, prevent device from sleeping & Data to be shared with third party & Pakistan & 500,000+ & No & lack of encryption. Data might be traveling in plaintext. Radius alert is not accurate. Doesn’t show patients infected with COVID-19 & 3.0.7 &	5.0 and up\\
\hline
COVID Symptom Tracker &    Android \& iOS    & Wifi connection information, full network access, audio settings, run at startup, prevent the device from sleeping &  Collects sensitive personal information such as DOB, name, gender, COVID-19 tests status, location, details of any treatment, email, phone number, IP address. Shared with universities, research centers, amazon web service, google analytic, etc.  & United Kingdom  & 500,000+    & N/A &  basic information related to COVID-19 symptoms, helps people take precautionary measures to self-isolate &0.14 &	5.0 and up \\
\hline
BC COVID-19 Support &    Android \& iOS & Location (approximate and precise), full network access, prevent the device from sleeping & Personal information collected for COVID-19 alerts and management, only used by Ministry of Health &    Canada    & 10,000+ & Yes &  Doesn’t update on a regular basis with current stats. No graph of active cases. Inaccurate and outdated information & 1.20.0 &	5.0 and up\\
\hline
OpenWHO: Knowledge for Health Emergencies &    Android \& iOS & Wifi connection, full network access, media/files, and storage, run at startup, prevent the device from sleeping & Requires name and email to create an account used for communications and the announcement of changes to the openWHO platform &    United States &    500,000+ &    No &    Language issue to some people. Gives out a certificate for completing the course, increases public health knowledge & 3.4 &	5.0 and up \\
%\hline
%COVID19 Care for Media &    Android \&iOS &    SMS, full network access, prevent the device from sleeping &    May collect sensitive data such as NIC, name, city, province and share this with third party    & Pakistan     & 10,000+    & Yes    &    Gives detailed information related to COVID-19 protection, symptoms & 8 &	6.0 and up\\
\hline
%Health Assessment PDMA &    Android \& iOS &    Location (approximate and precise), full network access  & Information accessed by Smart Asset Sindh Health, shared with third party     &    Pakistan &    10,000+    & No    & No encryption or algorithm is used for data protection. Too many bugs. Patchy functionality & 2.0.1	& 4.2 and up \\
%\hline
Pakistan's National Action Plan for COVID-19 &    Android    & This application requires no special permissions to run    & No information being shared    &     Pakistan &    50,000+    & Yes    &    Shares information related to COVID-19 and SOPs that government has launched for the safety of people & 1.1	& 5.0 and up\\
\hline
Health Monitoring PDMA    & Android &    Location (approximate and precise), receives data from internet and full network access    & Information will be accessed by Smart Asset Sindh Health, shared with third party &    Pakistan    & 1000+    & No &    Data information being sent in plaintext. No encryption or algorithm is used for data protection. Too many bugs. Doesn’t work efficiently. & 1.4	& 4.0.3 and up\\
\hline

%COVID-19 Screening Tool &    Apple    & -   &High privacy, data sharing between apple &    USA & - &    -    & \\            \hline
Canada COVID-19    & Android & Location (approximate and precise), full network access, prevent device from sleeping &    Personal data is collected by Health Canada only to support COVID-19.  &Canada    & 50,000+    & Yes & App doesn’t take into account pre-existing conditions. Will be much more effective if user can see map with active cases. & 4.0.0	& 5.0 and up\\
\hline

COVID-19  & Android &    Location, phone, media, storage, camera, microphone, wifi, device ID, call information, download files without notification, run at startup, prevent the device from sleeping  & May use personal information with third party &    Vietnam &    100,000+ &    N/A    &    App is only available for Veitnamese and not available in English. Very narrow coverage overall. Provides basic information. Only accessible in Veitnam & 1	& 4.4 and up\\
\hline

COVI    & Android &    Location, phone, wifi, device ID, call information, pair with Bluetooth devices, receive data from Internet, run at startup, prevent device from sleeping &    Information such as DOB, name, the account number is collected and shared with trusted third parties & Qatar &    10,000+ &    N/A & Only restricted for the people living in Qatar.  Doesn’t get updates. Provides basic information & 2.0.2.2	& 5.1 and up \\
\hline

COVID19 - DXB Smart App &    Android    & Microphone, camera, location, storage, calendar, Wifi connection, media, receive data from Internet, pair with Bluetooth devices, full network access, prevent device from sleeping, change audio settings & Sends personal information such as ID, name, DOB, email, geographical location to a third party & Dubai    & 1000+ & N/A &   Only restricted for the people living in Dubai. Some users reported experiencing network error whenever they open this app & 3.8 &	5.0 and up\\
\hline

Corona 360    & Android &    Location (approximate and precise), receives data from the internet, full network access, prevent the device from sleeping &    Collects personal and sensitive data of user but whenever the data is being used, the user is notified for the reason    & South Korea &    10+    Yes &    & Useful and multilingual solution for find Corona free locations & 2.2.2	& 4.3 and up \\
\hline
CoronaCheck    & Android \&iOS &    Full network access     & Will not share any information &    Pakistan    & 10,000+     & Yes    &    Gives detailed information related to COVID-19 protection, symptoms. Translate English to Vocal language. & 1.1	& 4.1 and up \\
\hline
Coronavirus Australia    & Android \&iOS    & Location (approximate and precise), receives data from internet and full network access &    Collects information but does not use it without asking from the user &    Australia    & 500,000+    &     N/A & App opens in the web browser which is clunky, the infection status is updated less often than the press releases, and is out of date later in the day. & 1.4.5	& 6.0 and up \\
\hline
NHS App    & Android \& iOS    & Location, phone, media, storage, camera, microphone, Wifi, device ID, call information, download files without notification, run at startup, prevent device from sleeping    & No specific information about sharing data with third parties &United Kingdom &    500,000+ &    N/A &    Requires personal details such as photo, name, DOB, NHS number. Requires 12 hours for the initial setup. Misleading/inaccurate information about compatible operating systems & 1.36.3 &	5.0 and up\\
\hline
Aarogya Setu &    Android    & Location (approximate and precise), receives data from internet and full network access &    Cannot access its privacy policy    & India    & 50,000,000+ &            N/A &  Location, network and Bluetooth visibility  required.   No proper tracking, no radius alert, bugs, doesn’t update cases. Takes a new location every time when accessed & 1.4.1	& 5.0 and up\\
\hline
%Apple COVID-19    & Apple    &    N/A & High privacy, data sharing with apple &     United States & N/A & N/A& N/A             \\ \hline

%HEALTH LYNKED COVID-19 Tracker &     iOS    & N/A &    High privacy, data sharing with apple &    United States & N/A & N/A& N/A                    \\ \hline

TraceTogether & Android \& iOS & Media, storage, receive data from Internet, pair wth Bluetooth devices, full internet access, prevent device from sleeping & Mobile number and anonymous ID are shared in a secure server and not available to be shared with Public &  Singapore & 500,000+ & N/A & Doesn't alert you to infected cases in your area. Drains battery pretty fast due to Bluetooth connection. & 2.2.0	& 5.1 and up\\ \hline

HaMagen    & Android \&iOS    & Device and app history, location, Wifi connection, full network access, prevent device from sleeping, change network connectivity & Cross-referencing location data with the corona patients & Israel & 1,000,000+  & No & Correlates overlaps only since installation. Should extract and use historical information. Data processed locally when a user opts against downloaded "Corona paths" & 2.2.6	&  5.0 and up\\ \hline

%Corona-Datenspende &    Android/ Apple &    Wifi connection and full network access & Collects personal data such as name, DOB, gender, location, IP address, however, implement security to protect this data &    Private    & 100,000+     & N/A & one cannot use the app without connecting to a fitness account and hence completely breaking the point of anonymity. Doesn't work with my Polar M430. Login to Fitbit not possible using latest Android on OnePlus7Pro - neither email/password (not possible to input credentials) not login with Google/Facebook (not yet supported). \\ \hline

Home Quarantine (Kwarantanna domowa) & Android \& iOS & Location, phone, media, storage, camera, microphone, wifi, device ID, call information, download files without notification, run at startup, prevent device from sleeping &    Collected data may be shared with third party &    Poland &   100,000+ &    N/A &GPS location is invalid. Cannot add a phone number as it gives away error. & 1.39.5	& 6.0 and up\\ \hline

%WHO Info &    Android \& iOS & Full network access, media, storage    & Collect personal data of the user for the record but do not share with third parties  &    Switzerland    & 100,000+     & Yes    & Very little data or charts. Not consistent with Worldometer. Mostly an aggregation of news feeds about WHO & 2.2.0	& 4.2 and up\\\hline

NHS 24 : COVID-19 &    Android \& iOS &    Full network access, receives data from internet prevent device from sleeping &    Collect personal data and share with third party &    UK    & 1000+        &    N/A &Not Compatible     Basic information only & 1.0.3	& 4.1 and up\\ \hline

Beat Covid Gibraltar &   Android &   view Wi-Fi connection,pair with Bluetooth devices, full network access &    No personal data will be stored or used &    Gibraltar    & 10000+        &    N/A &Developed only for Gibraltar, easy to use & 1.18	& 6 and up\\ \hline

EHTERAZ &   Android &   location data, phone access for calls, Photos / Media / Files,  full network access &   personal data will be stored or used &    Qatar  & 1,000,000+ &  n/A & additional authentication performed, some privacy flaws are identified & 9.02	& 6 and up\\ \hline

BeAware Bahrain &   Android &  require access to apps running, read calendar information, require access to location, media files, and storage, pair Bluetooth devices &   personal data will be stored or used &    Bahrain  & 100,000+ &    N/A &some privacy flaws are identified & 0.2.1	& 4.4 and up\\ \hline

Shlonik
&   Android &  record audio, access to running apps, require access to location, media files, and storage, pair Bluetooth devices, precise location information, full network access & Collects data and location information&    Kuwait  & 100,000+ &    N/A &some privacy flaws are identified & varies	& 4.4 and up\\
\hline

COVID Radar
&   Android/iOS & phone access, access to media files and storage, pair Bluetooth devices & Data is provided by users manually&    Netherlands  & 50,000+ &    N/A & N/A& 1.1.2	& 6 and up\\ \hline

Tawakkalna (Covid-19 KSA)&   Android & GPS Location, read the storage data, take pictures and make video, pair via Bluetooth device, full network access & Data is provided by users manually&    Saudi Arabia & 1,000,000+&    N/A & helps in imposing curfew & 1.7	& 6 and up\\ \hline

MySejahtera &   Android & precise location (GPS and network-based), call access, media \& storage access, camera access, full network access
 & Data is provided by users manually&   Malaysia & 1000K+ &    N/A & user also need to register through their website & 1.0.24	& 4 and up\\
\hline

\end{longtable}
\label{tab:analysis}

%\end{table*}
\twocolumn

\bibliographystyle{IEEEtran}
\bibliography{Reference.bib}

\begin{thebibliography}{10}
\providecommand{\url}[1]{#1}
\csname url@samestyle\endcsname
\providecommand{\newblock}{\relax}
\providecommand{\bibinfo}[2]{#2}
\providecommand{\BIBentrySTDinterwordspacing}{\spaceskip=0pt\relax}
\providecommand{\BIBentryALTinterwordstretchfactor}{4}
\providecommand{\BIBentryALTinterwordspacing}{\spaceskip=\fontdimen2\font plus
\BIBentryALTinterwordstretchfactor\fontdimen3\font minus
  \fontdimen4\font\relax}
\providecommand{\BIBforeignlanguage}[2]{{%
\expandafter\ifx\csname l@#1\endcsname\relax
\typeout{** WARNING: IEEEtran.bst: No hyphenation pattern has been}%
\typeout{** loaded for the language `#1'. Using the pattern for}%
\typeout{** the default language instead.}%
\else
\language=\csname l@#1\endcsname
\fi
#2}}
\providecommand{\BIBdecl}{\relax}
\BIBdecl

\bibitem{Ferrettieabb6936}
\BIBentryALTinterwordspacing
L.~Ferretti, C.~Wymant, M.~Kendall, L.~Zhao, A.~Nurtay, L.~Abeler-D{\"o}rner,
  M.~Parker, D.~Bonsall, and C.~Fraser, ``Quantifying sars-cov-2 transmission
  suggests epidemic control with digital contact tracing,'' \emph{Science},
  vol. 368, no. 6491, 2020. [Online]. Available:
  \url{https://science.sciencemag.org/content/368/6491/eabb6936}
\BIBentrySTDinterwordspacing

\bibitem{southkorea}
\BIBentryALTinterwordspacing
(2017) Test, trace, contain: how south korea flattened its coronavirus curve.
  [Online]. Available: \url{https://tinyurl.com/yabnmckr}
\BIBentrySTDinterwordspacing

\bibitem{southkore1}
\BIBentryALTinterwordspacing
(2017) How digital contact tracing slowed covid-19 in east asia. [Online].
  Available: \url{https://tinyurl.com/tc8z8wj}
\BIBentrySTDinterwordspacing

\bibitem{Troncoso2020}
C.~Troncoso, M.~Payer, J.-P. Hubaux, M.~Salath{\'e}, J.~R. Larus, E.~Bugnion,
  W.~Lueks, T.~Stadler, A.~Pyrgelis, D.~Antonioli, L.~Barman, S.~Chatel, K.~G.
  Paterson, S.~vCapkun, D.~Basin, J.~Beutel, D.~Jackson, M.~Roeschlin, P.~Leu,
  B.~Preneel, N.~P. Smart, A.~Abidin, S.~Gurses, M.~Veale, C.~J.~F. Cremers,
  M.~Backes, N.~O. Tippenhauer, R.~Binns, C.~Cattuto, A.~Barrat, D.~Fiore,
  M.~Barbosa, R.~Oliveira, and J.~C. Pereira, ``Decentralized
  privacy-preserving proximity tracing,'' \emph{ArXiv}, vol. abs/2005.12273,
  2020.

\bibitem{ga1}
A.~{Narayanan} and V.~{Shmatikov}, ``De-anonymizing social networks,'' in
  \emph{2009 30th IEEE Symposium on Security and Privacy}, 2009, pp. 173--187.

\bibitem{ga2}
------, ``Robust de-anonymization of large sparse datasets,'' in \emph{2008
  IEEE Symposium on Security and Privacy (sp 2008)}, 2008, pp. 111--125.

\bibitem{ga3}
\BIBentryALTinterwordspacing
K.~Sharad and G.~Danezis, ``An automated social graph de-anonymization
  technique,'' in \emph{Proceedings of the 13th Workshop on Privacy in the
  Electronic Society}, ser. WPES ’14.\hskip 1em plus 0.5em minus 0.4em\relax
  New York, NY, USA: Association for Computing Machinery, 2014, p. 47–58.
  [Online]. Available: \url{https://doi.org/10.1145/2665943.2665960}
\BIBentrySTDinterwordspacing

\bibitem{IND2020}
\BIBentryALTinterwordspacing
(2020) Coronavirus: Nhs contact tracing app needs 60\% take-up to be
  successful, expert warns. [Online]. Available:
  \url{https://tinyurl.com/y7u3mh7a}
\BIBentrySTDinterwordspacing

\bibitem{S1}
\BIBentryALTinterwordspacing
K.~W.~Y. Au, Y.~F. Zhou, Z.~Huang, and D.~Lie, ``Pscout: Analyzing the android
  permission specification,'' in \emph{Proceedings of the 2012 ACM Conference
  on Computer and Communications Security}, ser. CCS ’12.\hskip 1em plus
  0.5em minus 0.4em\relax New York, NY, USA: Association for Computing
  Machinery, 2012, p. 217–228. [Online]. Available:
  \url{https://doi.org/10.1145/2382196.2382222}
\BIBentrySTDinterwordspacing

\bibitem{S2}
W.~Enck, P.~Gilbert, B.-G. Chun, L.~P. Cox, J.~Jung, P.~McDaniel, and A.~N.
  Sheth, ``Taintdroid: An information-flow tracking system for realtime privacy
  monitoring on smartphones,'' in \emph{Proceedings of the 9th USENIX
  Conference on Operating Systems Design and Implementation}, ser.
  OSDI’10.\hskip 1em plus 0.5em minus 0.4em\relax USA: USENIX Association,
  2010, p. 393–407.

\bibitem{S3}
\BIBentryALTinterwordspacing
L.~Shi, J.~Fu, Z.~Guo, and J.~Ming, ``“jekyll and hyde” is risky:
  Shared-everything threat mitigation in dual-instance apps,'' in
  \emph{Proceedings of the 17th Annual International Conference on Mobile
  Systems, Applications, and Services}, ser. MobiSys ’19.\hskip 1em plus
  0.5em minus 0.4em\relax New York, NY, USA: Association for Computing
  Machinery, 2019, p. 222–235. [Online]. Available:
  \url{https://doi.org/10.1145/3307334.3326072}
\BIBentrySTDinterwordspacing

\bibitem{ct12}
T.~Altuwaiyan, M.~Hadian, and X.~Liang, ``Epic: Efficient privacy-preserving
  contact tracing for infection detection,'' \emph{2018 IEEE International
  Conference on Communications (ICC)}, pp. 1--6, 2018.

\bibitem{ct13}
L.~O. Danquah, N.~Hasham, M.~MacFarlane, F.~E. Conteh, F.~Momoh, A.~A. Tedesco,
  A.~Jambai, D.~A. Ross, and H.~A. Weiss, ``Use of a mobile application for
  ebola contact tracing and monitoring in northern sierra leone: a
  proof-of-concept study,'' \emph{BMC Infectious Diseases}, vol.~19, 2019.

\bibitem{ct14}
A.~Prasad and D.~Kotz, ``Enact: Encounter-based architecture for contact
  tracing,'' 06 2017, pp. 37--42.

\bibitem{ct15}
E.~Reddy, S.~Kumar, N.~Rollings, and R.~Chandra, ``Mobile application for
  dengue fever monitoring and tracking via gps: Case study for fiji,'' 03 2015.

\bibitem{ct16}
\BIBentryALTinterwordspacing
C.~Shahabi, L.~Fan, L.~Nocera, L.~Xiong, and M.~Li, ``Privacy-preserving
  inference of social relationships from location data: A vision paper,'' in
  \emph{Proceedings of the 23rd SIGSPATIAL International Conference on Advances
  in Geographic Information Systems}, ser. SIGSPATIAL ’15.\hskip 1em plus
  0.5em minus 0.4em\relax New York, NY, USA: Association for Computing
  Machinery, 2015. [Online]. Available:
  \url{https://doi.org/10.1145/2820783.2820880}
\BIBentrySTDinterwordspacing

\bibitem{ct17}
E.~Yoneki and J.~Crowcroft, ``Epimap: Towards quantifying contact networks for
  understanding epidemiology in developing countries,'' \emph{Ad Hoc Networks},
  vol.~13, p. 83–93, 02 2014.

\bibitem{ct11}
A.~Hekmati, G.~Ramachandran, and B.~Krishnamachari, ``Contain: Privacy-oriented
  contact tracing protocols for epidemics,'' 2020.

\bibitem{eu1}
\BIBentryALTinterwordspacing
(2017) General data protection regulation gdpr. [Online]. Available:
  \url{https://gdpr-info.eu/}
\BIBentrySTDinterwordspacing

\bibitem{eu2}
\BIBentryALTinterwordspacing
(2017) Mobile privacy principles promoting consumer privacy in the mobile
  ecosystem. [Online]. Available: \url{https://tinyurl.com/yaxy74wa}
\BIBentrySTDinterwordspacing

\bibitem{ICO}
\BIBentryALTinterwordspacing
(2017) Mobile privacy disclosures: Building trust through transparency: A
  federal trade commission staff report. [Online]. Available:
  \url{https://tinyurl.com/y765bgxj}
\BIBentrySTDinterwordspacing

\bibitem{appper}
A.~P. Felt, K.~Greenwood, and D.~Wagner, ``The effectiveness of application
  permissions,'' in \emph{Proceedings of the 2nd USENIX Conference on Web
  Application Development}, ser. WebApps’11.\hskip 1em plus 0.5em minus
  0.4em\relax USA: USENIX Association, 2011, p.~7.

\bibitem{per1}
\BIBentryALTinterwordspacing
(2017) Many popular android apps leak sensitive data, leaving millions of
  consumers at risk. [Online]. Available: \url{https://tinyurl.com/yb7hfjxr}
\BIBentrySTDinterwordspacing

\bibitem{per4}
\BIBentryALTinterwordspacing
(2017) Researchers spot thousands of android apps leaking user data through
  misconfigured firebase databases. [Online]. Available:
  \url{https://tinyurl.com/ybjdrcth}
\BIBentrySTDinterwordspacing

\bibitem{per2}
\BIBentryALTinterwordspacing
B.~Krishnamurthy and C.~E. Wills, ``On the leakage of personally identifiable
  information via online social networks,'' in \emph{Proceedings of the 2nd ACM
  Workshop on Online Social Networks}, ser. WOSN ’09.\hskip 1em plus 0.5em
  minus 0.4em\relax New York, NY, USA: Association for Computing Machinery,
  2009, p. 7–12. [Online]. Available:
  \url{https://doi.org/10.1145/1592665.1592668}
\BIBentrySTDinterwordspacing

\bibitem{per3}
C.~{Zuo}, Z.~{Lin}, and Y.~{Zhang}, ``Why does your data leak? uncovering the
  data leakage in cloud from mobile apps,'' in \emph{2019 IEEE Symposium on
  Security and Privacy (SP)}, 2019, pp. 1296--1310.

\bibitem{un1}
J.~Kang, D.~Kim, H.~Kim, and J.~H. Huh, ``Analyzing unnecessary permissions
  requested by android apps based on users' opinions,'' in \emph{Information
  Security Applications}, K.-H. Rhee and J.~H. Yi, Eds.\hskip 1em plus 0.5em
  minus 0.4em\relax Cham: Springer International Publishing, 2015, pp. 68--79.

\bibitem{un2}
\BIBentryALTinterwordspacing
S.~T. Peddinti, I.~Bilogrevic, N.~Taft, M.~Pelikan, U.~Erlingsson,
  P.~Anthonysamy, and G.~Hogben, ``Reducing permission requests in mobile
  apps,'' in \emph{Proceedings of the Internet Measurement Conference}, ser.
  IMC ’19.\hskip 1em plus 0.5em minus 0.4em\relax New York, NY, USA:
  Association for Computing Machinery, 2019, p. 259–266. [Online]. Available:
  \url{https://doi.org/10.1145/3355369.3355584}
\BIBentrySTDinterwordspacing

\bibitem{muhammad}
M.~{Ikram} and M.~A. {Kaafar}, ``A first look at mobile ad-blocking apps,'' in
  \emph{2017 IEEE 16th International Symposium on Network Computing and
  Applications (NCA)}, 2017, pp. 1--8.

\bibitem{muhammad1}
\BIBentryALTinterwordspacing
M.~Ikram, N.~Vallina-Rodriguez, S.~Seneviratne, M.~A. Kaafar, and V.~Paxson,
  ``An analysis of the privacy and security risks of android vpn
  permission-enabled apps,'' in \emph{Proceedings of the 2016 Internet
  Measurement Conference}, ser. IMC ’16.\hskip 1em plus 0.5em minus
  0.4em\relax New York, NY, USA: Association for Computing Machinery, 2016, p.
  349–364. [Online]. Available: \url{https://doi.org/10.1145/2987443.2987471}
\BIBentrySTDinterwordspacing

\bibitem{Liccardi}
\BIBentryALTinterwordspacing
I.~Liccardi, J.~Pato, and D.~J. Weitzner, ``Improving user choice through
  better mobile apps transparency and permissions analysis,'' \emph{Journal of
  Privacy and Confidentiality}, vol.~5, no.~2, Feb. 2014. [Online]. Available:
  \url{https://journalprivacyconfidentiality.org/index.php/jpc/article/view/630}
\BIBentrySTDinterwordspacing

\bibitem{berral}
\BIBentryALTinterwordspacing
D.~Barrera, H.~G. Kayacik, P.~C. van Oorschot, and A.~Somayaji, ``A methodology
  for empirical analysis of permission-based security models and its
  application to android,'' in \emph{Proceedings of the 17th ACM Conference on
  Computer and Communications Security}, ser. CCS ’10.\hskip 1em plus 0.5em
  minus 0.4em\relax New York, NY, USA: Association for Computing Machinery,
  2010, p. 73–84. [Online]. Available:
  \url{https://doi.org/10.1145/1866307.1866317}
\BIBentrySTDinterwordspacing

\bibitem{enck}
\BIBentryALTinterwordspacing
W.~Enck, M.~Ongtang, and P.~McDaniel, ``On lightweight mobile phone application
  certification,'' in \emph{Proceedings of the 16th ACM Conference on Computer
  and Communications Security}, ser. CCS ’09.\hskip 1em plus 0.5em minus
  0.4em\relax New York, NY, USA: Association for Computing Machinery, 2009, p.
  235–245. [Online]. Available: \url{https://doi.org/10.1145/1653662.1653691}
\BIBentrySTDinterwordspacing

\bibitem{pern}
\BIBentryALTinterwordspacing
P.~H. Chia, Y.~Yamamoto, and N.~Asokan, ``Is this app safe? a large scale study
  on application permissions and risk signals,'' in \emph{Proceedings of the
  21st International Conference on World Wide Web}, ser. WWW ’12.\hskip 1em
  plus 0.5em minus 0.4em\relax New York, NY, USA: Association for Computing
  Machinery, 2012, p. 311–320. [Online]. Available:
  \url{https://doi.org/10.1145/2187836.2187879}
\BIBentrySTDinterwordspacing

\bibitem{max}
\BIBentryALTinterwordspacing
M.~Van~Kleek, I.~Liccardi, R.~Binns, J.~Zhao, D.~J. Weitzner, and N.~Shadbolt,
  ``Better the devil you know: Exposing the data sharing practices of
  smartphone apps,'' in \emph{Proceedings of the 2017 CHI Conference on Human
  Factors in Computing Systems}, ser. CHI ’17.\hskip 1em plus 0.5em minus
  0.4em\relax New York, NY, USA: Association for Computing Machinery, 2017, p.
  5208–5220. [Online]. Available:
  \url{https://doi.org/10.1145/3025453.3025556}
\BIBentrySTDinterwordspacing

\bibitem{ct7}
T.~Yasaka, B.~Lehrich, and R.~Sahyouni, ``Peer-to-peer contact tracing:
  Development of a privacy-preserving smartphone app,'' 04 2020.

\bibitem{ross2}
\BIBentryALTinterwordspacing
(2017) Joint statement on contact tracing: Date 19th april 2020. [Online].
  Available: \url{https://tinyurl.com/ybm27jxx}
\BIBentrySTDinterwordspacing

\bibitem{ross1}
\BIBentryALTinterwordspacing
(2017) Contact tracing in the real world. [Online]. Available:
  \url{https://tinyurl.com/ty7sxo5}
\BIBentrySTDinterwordspacing

\bibitem{Cho2020ContactTM}
H.~Cho, D.~Ippolito, and Y.~W. Yu, ``Contact tracing mobile apps for covid-19:
  Privacy considerations and related trade-offs,'' \emph{ArXiv}, vol.
  abs/2003.11511, 2020.

\bibitem{manish}
M.~Shukla, A.~RajanM., S.~Lodha, G.~Shroff, and R.~Raskar, ``Privacy guidelines
  for contact tracing applications,'' \emph{ArXiv}, vol. abs/2004.13328, 2020.

\bibitem{ethics}
\BIBentryALTinterwordspacing
M.~J. Parker, C.~Fraser, L.~Abeler-D{\"o}rner, and D.~Bonsall, ``Ethics of
  instantaneous contact tracing using mobile phone apps in the control of the
  covid-19 pandemic,'' \emph{Journal of Medical Ethics}, 2020. [Online].
  Available: \url{https://tinyurl.com/ybojkjr8}
\BIBentrySTDinterwordspacing

\bibitem{comp}
\BIBentryALTinterwordspacing
(2017) Privacy \& pandemics: The role of mobile apps (chart). [Online].
  Available: \url{https://tinyurl.com/ycyw9tpk}
\BIBentrySTDinterwordspacing

\bibitem{cryptoeprint}
S.~Vaudenay, ``Analysis of dp3t,'' Cryptology ePrint Archive, Report 2020/399,
  2020, \url{https://eprint.iacr.org/2020/399}.

\bibitem{sun2020vetting}
R.~Sun, W.~Wang, M.~Xue, G.~Tyson, S.~Camtepe, and D.~Ranasinghe, ``Vetting
  security and privacy of global covid-19 contact tracing applications,'' 2020.

\bibitem{Yaron}
Y.~Gvili, ``Security analysis of the covid-19 contact tracing specifications by
  apple inc. and google inc.'' Cryptology ePrint Archive, Report 2020/428,
  2020, \url{https://eprint.iacr.org/2020/428}.

\bibitem{cdc}
\BIBentryALTinterwordspacing
(2017) Preliminary criteria for the evaluation of digital contact tracing tools
  for covid-19. [Online]. Available: \url{https://tinyurl.com/y7nroqlw}
\BIBentrySTDinterwordspacing

\bibitem{ct1}
H.~Cho, D.~Ippolito, and Y.~W. Yu, ``Contact tracing mobile apps for covid-19:
  Privacy considerations and related trade-offs,'' 2020.

\bibitem{ct9}
J.~A. Sacks, E.~Zehe, C.~Redick, A.~Bah, K.~Cowger, M.~Camara, A.~Diallo,
  A.~N.~I. Gigo, R.~S. Dhillon, and A.~Liu, ``Introduction of mobile health
  tools to support ebola surveillance and contact tracing in guinea,''
  \emph{Global Health: Science and Practice}, vol.~3, pp. 646 -- 659, 2015.

\bibitem{ct10}
\BIBentryALTinterwordspacing
B.~Armbruster and M.~Brandeau, ``{Contact tracing to control infectious
  disease: when enough is enough},'' \emph{Health Care Management Science},
  vol.~10, no.~4, pp. 341--355, December 2007. [Online]. Available:
  \url{https://tinyurl.com/y9yr4rry}
\BIBentrySTDinterwordspacing

\bibitem{RePEc}
\BIBentryALTinterwordspacing
B.~Zhang, S.~E. Kreps, and N.~McMurry, ``{Americans' perceptions of privacy and
  surveillance in the COVID-19 Pandemic},'' Center for Open Science, OSF
  Preprints 9wz3y, May 2020. [Online]. Available:
  \url{https://ideas.repec.org/p/osf/osfxxx/9wz3y.html}
\BIBentrySTDinterwordspacing

\bibitem{tech}
\BIBentryALTinterwordspacing
F.~Christophe, A.-D. Lucie, F.~Luca, P.~Michael, K.~, Michelle, and B.~. David,
  ``{Digital contact tracing: comparing the capabilities of centralised and
  decentralised data architectures to effectively suppress the COVID-19
  epidemic whilst maximising freedom of movement and maintaining},'' Center for
  Open Science, OSF Preprints 9wz3y, May 2020. [Online]. Available:
  \url{https://ideas.repec.org/p/osf/osfxxx/9wz3y.html}
\BIBentrySTDinterwordspacing

\bibitem{ct18}
``Privacy-preserving contact tracing,''
  \url{https://www.apple.com/covid19/contacttracing}, 2020, [Online; accessed
  29-May-2020].

\bibitem{ct19}
``Apple and google partner on covid-19 contact tracing technology,''
  \url{https://tinyurl.com/wfw9ojr}, 2020, [Online; accessed 29-May-2020].

\bibitem{ct21}
M.~Scudellari, ``Covid-19 digital contact tracing: Apple and google work
  together as mit tests validity,'' \url{https://tinyurl.com/y87sljrz}, 2020,
  [Online; accessed 29-May-2020].

\bibitem{ct20}
D.~Leprince-Ringuet, ``The world's first contact-tracing app using google and
  apple's api goes live,'' \url{https://tinyurl.com/y9on8x2u}, 2020, [Online;
  accessed 29-May-2020].

\bibitem{ct29}
``Apple releases new covid-19 app and website based on cdc guidance,''
  \url{https://tinyurl.com/ya9mhzc8}, 2020, [Online; accessed 29-May-2020].

\bibitem{ct30}
``Covid-19 screening tool,'' \url{https://www.apple.com/covid19/}, 2020,
  [Online; accessed 29-May-2020].

\bibitem{ct40}
``Healthlynked corp.'s covid-19 tracker no.1 most downloaded app in apple
  medical store for march,'' \url{https://tinyurl.com/yafve87b}, 2020, [Online;
  accessed 29-May-2020].

\bibitem{ct41}
``Healthlynked,'' \url{https://www.healthlynked.com/corona-virus-tracker/},
  2020, [Online; accessed 29-May-2020].

\bibitem{ct22}
M.~of~Information Technology \&~Telecommunication, ``Application developed to
  deal with corona virus,'' \url{https://tinyurl.com/y78gzc4v}, 2020, [Online;
  accessed 29-May-2020].

\bibitem{ct31}
``Canada covid-19,'' \url{https://www.thrive.health/covid19-collection-notice},
  2020, [Online; accessed 29-May-2020].

\bibitem{ct32}
``Covid-19,'' \url{https://api.tetvietaic.com/pages/policye}, 2020, [Online;
  accessed 29-May-2020].

\bibitem{ct34}
``Covid19 - dxb smart app,'' \url{http://csms.ae/PrivacyPolicy.html}, 2020,
  [Online; accessed 29-May-2020].

\bibitem{ct33}
``Covi,'' \url{https://www.droobihealth.com/user-app/privacy}, 2020, [Online;
  accessed 29-May-2020].

\bibitem{ct35}
``Corona360,'' \url{https://corona-360.com/}, 2020, [Online; accessed
  29-May-2020].

\bibitem{ct36}
``Coronacheck,'' \url{https://tinyurl.com/y8l4aq9z}, 2020, [Online; accessed
  29-May-2020].

\end{thebibliography}






% \end{thebibliography}

\end{document}